# Enhanced MAC Parameters to Support Hybrid Dynamic Prioritization in MANETs

Hannah Monisha J.
Head, Department of Computer Science,
Indira Gandhi College of Arts and Science,
Govt. of Puducherry, India.

Rhymend Uthariaraj V
Professor and Director,
Ramanujan Computing Centre,
Anna University, Chennai, India.

## ABSTRACT
Quality of Service (QoS) for MANETs becomes a necessity because of its applications in decisive situations such as battle fields, flood and earth quake. Users belonging to diverse hierarchical category demanding various levels of QoS use MANETs. Sometimes, even a low category user may need to send an urgent message in time critical applications. Hence providing prioritization based on user category and urgency of the message the user is sending becomes necessary. In this paper we propose Enhanced MAC parameters to support Hybrid Dynamic priority in MANETs(H-MAC). It combines both prioritizations based on user categorization and dynamic exigency. Order Statistics is used to implement dynamic priority. We propose dynamic TXOP, Proportional AIFS and Proportional dynamic Backoff timers based on weights and collision, to avoid packet dropping and starvation of lower priorities. The model is simulated in ns2. We compare our results with IEEE 802.11e and show that, 16% more throughput is achieved by H-MAC during extensive collision. We also observe that starvation and packet drops are reduced with proportionate bandwidth sharing compared to the existing model.

## 1. INTRODUCTION
A MANET is a transitory network formed with a group of autonomous, on-the-fly mobile devices. Their topology is dynamic and there is no central coordinator. The nodes can self-organize and self-regulate. All the devices in a MANET function as a router for forwarding packets and at the same time send and receive packets. The applications of MANET are numerous. They can be used in a Classroom, Home, Business meetings and critical scenarios such as flood or earth quake.

The different type of data that flows in a MANET varies from voice and video to messages. Generally the real-time traffic such as voice and video require enhanced services than the others. Hence priority for resource allocation in any network is generally given to the real-time traffic. In MANETs, because of its application in critical scenarios, in addition to real-time traffic, emergency messages play a very crucial role[1]. These emergency messages require expedited and guaranteed transmission. Further the category of the user sending the message is also important because of the hierarchical organizational structure[2]. Hence, priority for accessing the bandwidth has to be given to the emergency non-real-time traffic based on the user category.

In this paper we propose a Hybrid dynamic priority algorithm which combines the static priority based on user profile and dynamic priority based on the urgency of the packet, which is estimated by the parameters, lifetime of the packet and the number of hops it has to traverse to reach the destination. Based on this we calculate the urgency index. An improved proportional share queue is used, where dequeuing is based on user defined weights, the percentage of queue length and collision rate. We further extend prioritization to the MAC layer with proportional share Arbitrary Inter frame Space (AIFS), dynamic Contention Window (CW) sizes and differentiated transmission opportunity limits (TXOP$_{limits}$). The model H-MAC is simulated in ns2. We compare our model with IEEE 802.11e. Results show that H-MAC performs better in terms of decreased packet drops, increased system throughput and starvation is avoided even during extensive collision in the network.

The rest of the paper is organized as follows. Section 2 presents Review of literature, Section 3 explains the proposed model, Section 4 gives Analysis of complexity, Section 5 details the Simulation results and Section 6 gives Conclusion and Future directions.

## 2. REVIEW OF LITERATURE
QoS in MANET though vital, is complex to accomplish because of its frequently changing topology, unreliable routes and power constraints. Researchers have ventured in finding various solutions for providing QoS for wired networks. They cannot be directly ported to MANET because of its unique characteristics. Since every node acts as a source, destination and a router, we need to realize QoS at every hop to attain overall QoS.

Though IEEE 802.11 DCF[3] does not support prioritization, IEEE 802.11e[4] EDCF supports service differentiation at the MAC layer. It supports four access categories (AC) where voice is given the highest priority followed by video, best effort and background. Differentiation is achieved at various stages such as arbitrary inter frame space (AIFS) and backoff. The AIFS is a prioritized static entity which is assigned for the four ACs. Lower the AIFS higher the priority. Backoff is set based on the CW size. CW is a static entity with variable window sizes varying between CWmin and CWmax for the various ACs. Lower the CW size, higher the priority. Standard transmission opportunity (TXOP) is new to IEEE 802.11e, which allows a burst of packets, till the TXOP$_{limit}$, which is assigned statically. The limitations of this protocol are discussed by various researchers. We have summarized a few. The small CW size assigned to high priority traffic leads to internal collision and thus increased packet drops especially when the high priority traffic is more[5]. Hence Adaptive CW sizes would be preferable. When the network is loaded and experiences high collision, static TXOP$_{limit}$ is inadequate. This is because majority of the TXOP duration would be spent on retransmissions[6]. Hence adjustable TXOP would be desired. Starvation of low priority traffic is experienced when the high priority traffic dominates the network. This is because of the priority queue scheduling model, where the low priority queue is serviced only when the higher priority queues are empty[5]. Hence hybrid scheduling algorithms have to be modeled. Sometimes, even if the load on the network is low, with only low priority traffic, the throughput of low priority traffic is low because of the static and low TXOP assigned to them. Further EDCF assigns priority statically to voice, video, data and background traffic. Priority is not given for non-real-time urgent messages, which has to be overcome.

Multihop latency aware scheduling (MLA)[7] proposes relative weights based on Lifetime-distance factor. It considers wireless fixed networks hence the distance does not change randomly. It is difficult to implement in MANET because of its mobility. The distance between source and destination node keep varying. Moreover, scheduling or MAC protocols are not proposed Adaptive delay threshold priority queuing (ADTPQ)[8] which is proposed for mobile broadband, assigns priority based on delay threshold. Delay threshold is calculated based on the cumulative system outrage. Adapting this as such for MANET is not effective because here there is no fixed set of users as in broadband services. Secondly, they propose one threshold for all type of traffic classes.

TXOP is calculated based on mean data rate, packet size and channel data rate by [6]. Drawback is that, if all the above mentioned parameters are constant, no prioritization will be achieved between the ACs. Further they do not consider network conditions. Adaptive EDCF[9] updates the contention window based on the average collision rate experienced by the station. The limitation is that, the multiplicative factor becomes constant, when collision increases. To overcome this [10] proposes Linear Adaptation(LA) and Hybrid Adaptation(HA) algorithm. In LA, only CWmin is updated. CWmax is not updated. This leads to static CW size which further increases collision. The problem of small window size is experienced as in IEEE 802.11e. In [11] we proposed dynamic scheduling. We have not considered Contention free burst(CFB) and proportional bandwidth sharing which would further reduce starvation and increase throughput.

To overcome certain limitations discussed above, we propose hybrid dynamic priority based on user category and urgency, Novel scheduling algorithm to overcome the drawbacks of priority scheduling and dynamic, proportionally differentiated MAC parameters to avoid packet drops and starvation.

## 3. PROPOSED MODEL – H-MAC
### 3.1. Classification of users
We meet the QoS requirements of users by classifying the users as High Profiled user(HP) with good QoS, Medium Profiled user(MP) with moderate QoS and Low Profiled user(LP) with Best effort service based on their profiles similar to [12]. To give importance to the urgent messages, we assign one more category Urgent Priority(UP) similar to [11]. Every user in the MANET is assigned a Static Priority Index (SPI) according to the classification(0-UP,1-HP, 2-MP, 3-LP). We assign proportional weights to different classes of users based on their hierarchical categorization such that, $w_{max} > w_0 > w_1 > w_2 > w_3 > 0$. $w_{max}$ is the maximum weight that can be assigned, $w_0$ is the weight of the urgent packets and $w_1$, $w_2$ and $w_3$ are the weights assigned for HP, MP and LP users. We reduce the ratio of weights based on the weight of LP for further calculations. When a packet is generated at source, we add two additional fields to the header of every packet to store the SPI and the Dynamic Urgency Index(DUI) calculated based on dynamic priority. We add a QoS flag(QF) to the header. The value of QF is set to 0 for packets generated from non QoS stations. If the user requests for QoS, then QF is set to 1.

### 3.2 Dynamic Urgency Index
To dynamically discriminate the packets, we calculate the urgency of the packet adapted from [11]. Urgency is determined dynamically at every hop with the parameters lifetime and hops. Lower the life time, more the number of hops, higher the urgency. To determine the urgency, we set delay threshold values. The initial delay threshold for a packet is calculated as the lifetime(LT) divided by the number of hops(H)[7]. This is considered as the initial delay threshold for a packet. The Local Delay (LD) at every node is calculated as the difference between the arrival time of the packet at the previous node ($a_{i-1}$) and the arrival time of the packet at the current node ($a_i$) as in equation(1). The generation time of the packet is assumed to be $a_0$.

$$LD_i = a_{i-1} - a_i \quad 1)$$

Now we find the Cumulative Delay because, if a node misses its local deadline at one hop, it may compensate at the next hop. Cumulative delay (CD) at node i is calculated as in equation(2).

$$CD_i = \sum_{j=1}^{i-1} LD_j \quad (2)$$

The packet can reach its destination on time only if it is able to maintain its delay threshold. We dynamically calculate the delay threshold at node $i$ using equation(3).

$$\tau_i = \frac{LT - CD_i}{H_i} \quad (3)$$

Where, $H_i$ is the remaining hop count from node $i$ to destination and the remaining lifetime is calculated by subtracting the cumulative delay CD from LT. Figure.1 explains this.

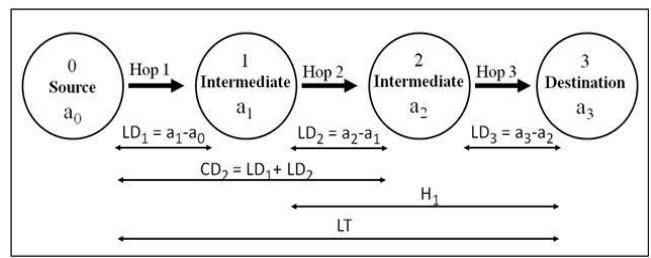

Figure1: Multihop MANET Model

If the local delay experienced at a node is less than the threshold $\tau_i$ then, a packet can reach on time. If the delay experienced at a node is greater than $\tau_i$ then the packet cannot reach on time. If the probability of a packet not reaching before its lifetime is more, then the packet is given higher priority. The probability $p$ of a packet reaching its destination before its lifetime i.e probability of success is calculated as the number of success divided by number of hops.

Lower the probability of success, higher should be the priority. We map probability and threshold to priority as in equation (5)

$DUI = \tau_i * p \quad (4)$

## 3.3. Hybrid Priority Scheduling

We consider SPI and DUI for scheduling. At every node, we maintain three separate queues based on user priority, such as the HP($Q_{HP}$), MP($Q_{MP}$) and LP($Q_{LP}$) users and one queue for urgent packets($Q_{UP}$) belonging to any of the user profiles. The packets are enqueued according to their classes in their respective queues. If QF=1, the urgent packets are enqueued in the urgent queue. The queues are ordered based on DUI. We implement proportional bandwidth sharing for all the queues including the urgent queue.

### 3.3.1 Modeling Scheduling with Order Statistics

We model our queues such that, at every TXOP, it selects the packet from the queue whose head of line(HOL) packet has the minimum DUI, for transmission. Since this method of selecting the minimal entity from the queue, relates to order statistics, our queue can be modeled using order statistics.

Assume that we have a basic random experiment, and that DUI of a packet which is a real-valued random variable for the experiment with distribution function F and probability density function f. We consider *n* independent replications of the basic experiment to generate a random sample X=($X_1$, $X_2$, …, $X_n$) which are considered as the packets waiting in the queue of size *n*. This is a sequence of independent random variables, each with the distribution of X. Let $Y_1$, $Y_2$,…$Y_n$ be the packets in the queue after ordering such that:$Y_1$=min{$X_1,X_2,…,X_n$}, $Y_n$=max{$X_1,X_2,…,X_n$} and $Y_1<Y_2<…Y_{k-1}< Y_k…<Y_n$. The packet $Y_k$ in the queue is called the kth order statistics[13]. The next step is to derive the distribution of the order statistics.

The Bernoulli trials are independent trials, each with two possible outcomes such as success and failure. They have the same probabilities from trial to trial. The probability of success p on a trial is the basic parameter of the process. The number of successes in *n* Bernoulli trials has the binomial distribution with parameters *n* and p, which has probability density function as in equation(5)

$$K \mapsto \binom{n}{k} p^k (1-p)^{n-k}, k \in \{0,1,...,n\} \quad (5)$$

Let $N_x$ be the number of sample variables less than or equal to x.

$$N_x = \sum_{i=1}^{n} 1(x_i \leq x), \ x \in \mathbb{R} \quad (6)$$

$N_x$ has the binomial distribution with parameters n and F(x) for each x∈ℝ. Now let $F_k$ denote the distribution function of the kth order statistic X(k). X(k) ≤ x if and only if $N_x$ ≥ k for x∈ℝ and k∈{1,2,…,n}.

From equations (5, 6), the distribution function of X(k) is given as

$$F_k(x) = \sum_{j=k}^{n} \binom{n}{j} [F(x)]^j [1-F(x)]^{n-j}, \ x \in \mathbb{R} \quad (7)$$

In particular, the cumulative distribution function (cdf) of the maximum and minimum of the variables of $X_{n:n}$ and $X_{1:n}$ is given in equation(8).

$$F_{n:n}(x) = F(x)^n$$

$$F_{1:n}(x) = 1 - [1 - F(x)]^n , \ x \in \mathbb{R} \quad (8)$$

Where $F_{1:n}(x) \leq F_{2:n}(x) \leq \cdots \leq F_{n:n}(x)$, because F is a non-decreasing function each of which is U(0,1)[13]. The objective of this research paper is to minimize the delay of urgent packets. From the derivations above, it can be observed the probability of delay would be in the increasing order of urgency and the delay of the most urgent packet will be minimal or zero according to first order statistics in equation(8). Using equation(8), the performance measures such as waiting time in the queue can be computed as in [15].

Since all the queues are internally ordered according to DUI at scheduling, the output would be an ordered list where, ordering is based on the DUI and User profile based priority. The lifetime of the packets can be adjusted so that we can enhance EDCF to support profile and traffic based priority. For example if the lifetime of the real-time traffic is assigned to be small, then subsequently it will be given a higher priority. Similarly, if the lifetime of an urgent data packet is small, it takes the higher priority. Thus this scheduling model can be used to design any kind of priority because it is ordered and analytically modeled.

The next step in scheduling is to decide the order in which the queue will be serviced. In IEEE 802.11e every queue in the node acts as a virtual node and contends for channel. At every contention, the highest priority queue acquires the channel first followed by the low priority ones similar to priority queuing. This is achieved by making the packets wait for differentiated backoff timers. The queue with its backoff at 0, wins the internal contention. Since the backoff timers of high priority ones are always small, the probability of it winning the contention is more. Thus the packets in lower priority queues are always starved.

### 3.3.2 Dynamic Proportional Bandwidth Sharing

To overcome starvation among low priority nodes, we proportionally access the queues. We decide the access ratios based on their proportional weights, queue length and collision rate of each class of packet. Here weights are constant but the queue length and collision rate are dynamic. The percentage of Queue length is calculated as in equation(9) and the collision rate is calculated using equation(10).

$$x_i = \frac{ql_i}{\sum_{j=0}^{3} ql_j} \quad (9)$$

Where, $ql_i$ and $x_i$ denote the queue length and the percentage of packets waiting in their respective $Q_{UP}$, $Q_{HP}$, $Q_{MP}$ and $Q_{LP}$. $x_i$ is updated whenever a packet is enqueued.

$$P_{col}^j[i] = (1 - \alpha) * P_{curr}^j[i] + \alpha * P_{col}^{j-1}[i], \ i=0 \ to \ 3 \quad (10)$$

Where $P_{col}^j[i]$ is the average packet collision ratio which varies between 0 and 1 adapted from [9]. $P_{curr}^j[i]$ is the current collision rate calculated during the j[th] update period as the number of collisions in class i divided by number of transmissions in class i. $\alpha$ is a smoothing factor in the range 0 and 1. To avoid stale collision rate, $P_{col}[i]$ is updated dynamically at equal time interval $T_{up}^j$. When the percentage of queue length of low priority packets is more or the collision ratio of low priority packets are more, the access ratio of High Priority packets may fall below the low priority ones leading to priority reversal. To prevent this, when the percentage of Higher Priority packets and collisions falls below the average percentage, we maintain a minimum bandwidth for the high priority traffic. Thus the percentage of Higher Priority packets are maintained at an average such that, *Av=100/4*. Thus at any point of time, for any random data flow, the priority hierarchy of packets is ensured. The following algorithm(1) calculates the Access ratio ($AR_i$) for each class based on their respective Queue length, Collision and weights.

**Algorithm-1 Calculation of Access Ratio**

**Step 1:** $cx_i = (x_i * (1 + P_{col}^j[i])) * 100$, $i=0$ to $3$

**Step 2:** *If $(cx_0<Av)$ then $cx_0 = Av$*

**Step 3:** *If $((cx_1>cx_0)$ or $(cx_1<cx_2)$ or $(cx_1<cx_3))$ then*

$cx_1 = Av$

**Step 4:** *If $((cx_2>x_0)$ or $(cx_2>x_1)$ or $(cx_2<cx_3))$ then $cx_2 = Av$*

**Step 5:** *If $((cx_3>cx_0)$ or $(cx_3>cx_1)$ or $(cx_3>cx_2))$ then $cx_3 = Av$*

**Step 6:** $AR_i = w_i * cx_i$, $i=0$ to $3$

## 3.4. Dynamic IEEE 802.11e MAC parameters

IEEE 802.11e is the MAC protocol widely used to achieve differentiated services at the MAC Layer. It uses four parameters to achieve differentiation. They are AIFS[i], CWmin[i], CWmax[i] and TXOP$_{limit}$ [i], where i represent various service classes. TXOP$_{limit}$ is a contention free burst, which specifies a time limit to dequeue the packets from the queue. TXOP is set long enough for a burst of a MSDU limited to a static parameter that is predefined, such that the TXOP of UP>HP>MP>LP. This static nature degrades the performance of the high priority traffic at two instances: 1) When the size of the queue is large, the high priority queues suffer packet dropping due to short and static TXOP. 2) When the collision on the network increases, the packets have to be retransmitted leading to less packet delivery ratio during the TXOP. To overcome these problems, we compute TXOP$_{limit}$ based on $AR_i$, which gives the number of packets to be dequeued based on weights, queue length and average collision ratio of each class. If there are no packets waiting in the higher priority queues, then the *TXOP$_{limit}$* value of the higher priority node is assigned to the lower priority ones, to increase the throughput of the low priority nodes. For example, if x$_0$=0 then the *TXOP$_{limit}$* of UP is assigned to HP. Similarly, if x$_0$=x$_1$=0 then *TXOP$_{limit}$* of UP is assigned to MP and so on. The following algorithm (2) achieves this.

**Algorithm-2 Calculation of TXOP**

**Step 1:** *TXOP$_{limit}$ [i] = (AR$_i$ * Tr) - SIFS, i=0 to 3*

**Step 2:** *If $x_{i-1}=0$ then TXOP$_{limit}$ [i]= TXOP$_{limit}$ [i-1], i=1 to 3*

Where *Tr* is the time required to transmit one data packet which is calculated as in equation(11)

$Tr = T_{data} + 2 * SIFS + T_{ACK}$ (11)

Here, $T_{data}$ is the time required to transmit one data packet based on MAC protocol Data Unit( MPDU), SIFS is the short inter frame space and $T_{ACK}$ is the time required to transmit ACK, which are PHY dependent. $AR_i$ and *TXOP$_{limit}$[i]* are updated after every burst. *Tr* is generally a static value, if the data is of Constant Bit Rate (CBR). This enhances the setting of more realistic and dynamic value for TXOP$_{limit}$. Hence the urgent queue which has the highest priority will have the maximum transmission opportunity, followed by HP, MP and LP queues. The low priority queues will not suffer starvation because of proportional sharing. In IEEE 802.11e RTS/CTS exchange is done optionally before a TXOP. In a MANET due to its mobility and ad hoc nature, RTS/CTS need to be exchanged whenever the destination varies, to ensure reliable transmission. To avoid complexity, we limit the TXOP to a burst of MSDU limited to the dynamic TXOP defined by us in algorithm(2).

Priority in granting a TXOP to a queue is achieved through backoff timers. There are two waiting stages during contention, the Arbitrary Inter Frame Space(AIFS) and the Back-off stage. IEEE 802.11e calculates AIFSN based on equation(12).

$AIFSN_i = SIFS + AIFSN_i * slot\text{-}time;$ $i=0$ to $3$ (12)

Where, AIFSN$_i$ is the static Arbitrary Inter Frame Space Number of class i. We propose to differentiate the AIFSN proportionally based on the weights[13]. It is calculated using algorithm(3) and then substituting in equation(12). The higher the weight assigned to a node, the lower will be the AIFSN and thus the priority will be higher. The AIFSN of UP<HP<MP<LP. Further, when the higher priority queue is empty, we assign the AIFSN of higher priority to lower priority nodes to decrease their delay.

**Algorithm-3 Calculation of AIFSN**

**Step 1:** *If $x_{i-1}=0$ then $AIFSN_i= AIFSN_{i-1}$, i=1 to 3*

**Step 2:** $AIFSN_i = integer\left(\frac{\sum_{j=0}^{3} w_j}{w_i}\right)$ ; $i=0$ to $3$

The next stage is the Back-off stage. IEEE 802.11e EDCA uses prioritized contention window sizes. Drawback is that the values are static. The shorter window sizes, may lead to increased packet drops because at every collision the contention window doubles. Hence we propose dynamic window sizes based on the collision by improving[9]. IEEE 802.11e uses a random access mechanism, where a node selects a backoff value based on the equation(13).

$Backoff[i] = integer(2^k * CW[i] * slot\text{-}time), i=0$ to $3$ (13)

Where CW[i] is a random integer value uniformly taking values in the range(0, CW[i]) inclusive. The initial value of CW[i] is set to CWmin[i]. At every collision, the CW doubles and the maximum value it can take is CWmax[i]. *k* is the number of attempts made for transmission. IEEE 802.11e uses very small CW for Higher priority. To overcome the limitation of small window size during collision, differentiated contention window sizes are proposed based on the collisions calculated by P$_{col}$[i] as in equation(10). We find the average collision rate (A$_{col}$) by summing the individual collision rates divided by number of classes. This is done because collision of any packet in the network will lead to further collisions with the same CW. To set the CW based on collision, we assign a collision threshold T$_{col}$ based on the collision tolerance. T$_{col}$ is carefully chosen in such a way that the CW size increases only when there is a need. Study of setting an optimal T$_{col}$ is out of scope of this paper. It will be taken up in the forthcoming study. If A$_{col}$ is greater than T$_{col}$ then we grow the size of the CW to avoid further collision and packet drop. If the collision rate is within the threshold, then the CW value is not altered. Further, if there is no packet waiting in the higher priority queues, then the lower priority ones are assigned small CW sizes, to increase the throughput of the low priority nodes. For example, if x$_0$=0 then, the CWmin and CWmax of UP are assigned to HP. Similarly, if x$_0$=x$_1$=0 then, the CWmin and CWmax of UP is assigned to MP and so on. The following algorithm(4) achieves this.

**Algorithm-4 Calculation of CWmin and CWmax**

| Step 1: | If $(x_{i-1}=0)$ then<br>$\{ CWmin_{new}[i] = CWmin_{old}[i-1],$<br>$CWmax_{new}[i] = CWmax_{old}[i-1]$<br>$\}$ $i=1$ to $3$ |
|---|---|
| Step 2: | if $(A_{col} \geq T_{col}])$ then<br>$\{$ $\{CWmax_{new}[i] = 2 * (CWmax_{old}[i] - CWmin_{old}[i])$<br>$CWmin_{new}[i+1] = CWmax_{new}[i]$<br>$\}$ $i=0$ to $2$<br>$CWmin_{new}[0] = CWmin_{old}[0]$<br>$CWmax_{new}[3] = CWmax_{old}[3]$<br>$\}$<br>Else<br>$\{$ $CWmin_{new}[i] = CWmin_{old}[i],$<br>$CWmax_{new}[i] = CWmax_{old}[i]$<br>$\}$ $i=0$ to $3$ |

$CWmin_{old}[i]$ and $CWmax_{old}[i]$ are the static values set in IEEE 802.11e. $CWmin_{new}[i]$ and $CWmax_{new}[i]$ are the dynamically assigned values. The CWmin and CWmax do not keep growing continuously at every collision. They grow only when the collision rate becomes greater than the threshold and are maintained at default when the collision is within the threshold.

To prioritize during retransmission, we modify equation(13) as equation(14) based on priority factor(PF).

*Backoff[i] = integer(PF[i]$^{2+i+k}$ * CW[i] * slot-time), i=0 to(14)*

We calculate PF[i] proportionate to the weights $w_i$. The lower the PF[i], lower will be the waiting time. Hence, PF[i] should be such that $0 < PF[0] < PF[1] < PF[2] < PF[3] < 1$. The following equation(15) calculates the PF[i] for UP, HP, MP, and LP proportional to their weights[12].

$P[i] = 1 - \dfrac{w_i}{\sum_{j=0}^{3} w_j}$ ; $i=0$ to $3$     (15)

This ensures prioritization during retransmissions.

According to the scheduling algorithm we proposed, the next TXOP would be granted to the queue whose HOL packet has the lowest DUI to enable transmission of urgent packets. If there is contention because of equal DUI, then it is resolved by choosing the lowest SPI.

## 4. ANALYSIS OF COMPLEXITY
The complexity of an algorithm is determined based on time and space complexity. In this section we analyze the complexity of H-MAC. The time complexity quantifies the time taken by the algorithm at runtime. It is generally estimated by counting the number of basic machine instructions such as add, subtract, multiply, divide, comparison and assignment. We calculate these basic instructions in our formulae and algorithms to find the time complexity of our model. The number of basic operations is considered a constant. The updation of equations(9,10) is done at a time interval $T_{up}$. If we consider 'n' as the number of times these equations are calculated, then the time complexity of our algorithm is linear time complexity O(n). If we consider that the data structure used to represent a queue is linked list, then the complexity of inserting an element in an ordered list using binary search is only O(log n). We use order statistics to order the four list and the complexity becomes 4*O(log n). Since this is done at every update 'm', the time complexity of our algorithm can be written as O(m log n). The space complexity is the storage required for execution of the algorithm. This includes all the temporary and permanent storage space required by the algorithm. The total space required by our algorithm is a constant with 'n' number of data in the queues. Thus the space complexity of our algorithm can be written as O(n). The time and space complexity of IEEE 802.11e is calculated as O(n).

## 5. SIMULATION AND RESULTS
Similar to wired networks, QoS in MANET can be measured in terms of throughput, delay, packet loss, jitter, packet delivery ratio etc. We implemented our proposed model and IEEE 802.11e in ns2. The test network included 5 to 100 nodes each assigned priorities such as HP, MP and LP randomly. The transmission range of each node is defined as 250m and the bandwidth of the channel is 2 Mbps. For the purpose of simulation, we have assigned the weights for UP, HP, MP, LP as $w_0=4$, $w_1=3$, $w_2=2$, $w_3=1$. We assume that the packets arrive at a Poisson distribution λ, and the service time at the queue is denoted by μ. Then the parameter ρ= $\dfrac{\lambda}{\mu}$ gives the traffic intensity or the congestion in the network. Other important parameters that favor the urgency of the packets are the hop count and lifetime of a packet. Given any source and destination node, we randomly set the lifetime of a packet such that it is uniformly distributed between hop count and maximum lifetime 20. Lower lifetime is allotted to packets requesting high urgency. We consider the hop count of a packet randomly between 1 and 10 adapted from [7]. $T_{up}$ is another important parameter that determines the period of updation. Too small a period will result in computation overhead and too long a period will be give stale report on the network. Hence we set $T_{up}$ to 5000 time-slot and α as 0.8 as in [9] which gives an appropriate tradeoff between goodput and delay. For simulation purpose, we set $T_{col}$, the collision threshold to 0.5 based on [10]. Three different scenarios were simulated altering the traffic conditions to study the performance of the model.

*Scenario 1* - Traffic has equal number of UP, HP, MP and LP packets in the queue. The collision rate is below 0.5.

*Scenario 2* – Traffic has equal number of UP, HP, MP and LP packets in the queue. The collision rate is above 0.5.

Multiple simulations were done and results were averaged. The QoS parameters for every scenario were recorded and analyzed.

### 5.1. Throughput
Throughput is calculated as the total number of bits received at the destination divided by the total transmission time. Figure 2, Shows the throughput results of Scenario I. It compares the throughput of IEEE 802.11e and H-MAC when the Collision rate is less than 0.5. It is observed that throughput of IEEE 802.11e and H-MAC shows performance such that throughput of UP>HP>MP>LP. The performance of IEEE 802.11e for UP is marginally better, when the number of nodes is less because, it follows priority queue scheduling. Throughput of H-MAC is observed to be better even when the number of nodes in the network increases. This is because of the Dynamic proportional bandwidth sharing approach that is used to fairly share bandwidth.

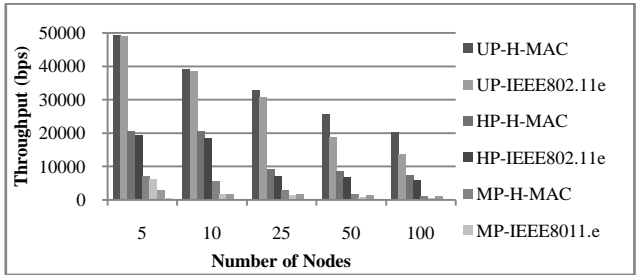

Figure 2: Comparative Throughput when Acol<Tcol

Figure 3, Shows the throughput results of Scenario II. It compares the throughput of IEEE 802.11e and H-MAC when the Collision rate is greater than 0.5. It is observed that throughput of IEEE 802.11e and H-MAC shows performance such that throughput of UP>HP>MP>LP. With IEEE 802.11e, though differentiation is maintained, when the number of nodes increases, collision increases, and the performance of LP degrades very badly. When the number of nodes in the network is 50 and 100, it observed that the LP nodes are completely starved. This is because of their priority scheduling algorithm. Throughput of H-MAC shows 16% improvement even when the number of nodes in the network increases and the collision is above threshold. This is because of the dynamically varying MAC parameters such as contention window size, AIFSN, and TXOP$_{limit}$ based on the channel condition and their weights.

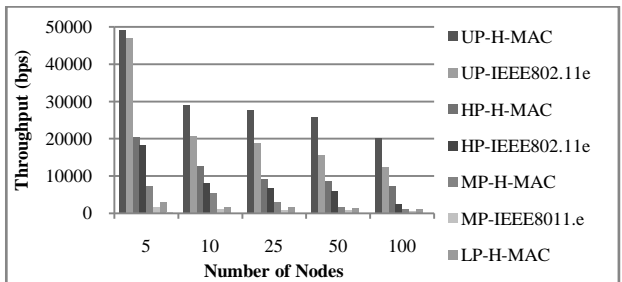

Figure 3: Comparative Throughput when Acol>=Tcol

## 5.2. Packet Delivery Ratio

Packet delivery ratio (PDR) is calculated as the ratio of the data packets delivered to the destinations to those generated by the CBR sources. Figure 4, shows the packet delivery ratio of both IEEE 802.11e and H-MAC for all classes of users. It can be observed that generally PDR decreases with the increase in the number of nodes in the network. Performance of LP is poor in IEEE 802.11e compared to LP-H-MAC because of the priority queue scheduling and large CW size. For UP and HP, when the congestion and collision is less in the network, packet loss is less. When number of node increases, packet loss increases even for UP and HP, because of the small contention window size and static TXOP$_{limit}$. Results show that, H-MAC performs better even during collision. Since we prioritize based on DUI which is calculated based on Lifetimes, packet loss has significantly reduced. Further packet loss is reduced because of the dynamic CW and TXOP$_{limit}$ which were set based on collision. This ensures the required packet transmission rate even at the time of collision.

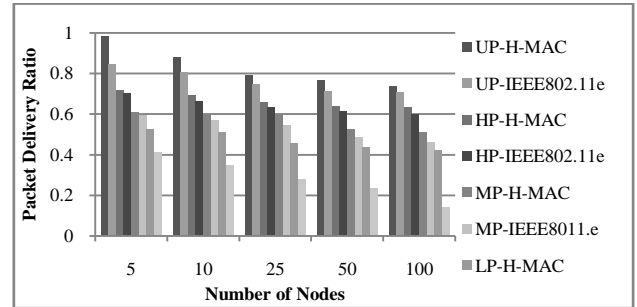

Figure.4. Comparative Packet Delivery Ratio

## 6. CONCLUSION

The objective of this research paper is to achieve differentiated services based on user profile and urgency of the packet and ensure fairness among all the priorities by avoiding starvation of Low Priority packets even during collision. In this paper we propose Hybrid priority Scheduling, Dynamic Proportional bandwidth Sharing and Enhanced dynamic MAC parameters to avoid packet dropping and starvation. Results show that our model gives 16% more average throughput during collision. We observe that starvation is reduced with proportionate shares and dynamic setting of AIFSN, CW and TXOP$_{limit}$ to a great extent compared to the existing model. Prioritization is highly influenced by Lifetime of a packet. As a future work, we plan to derive a method to calculate optimal lifetime for packets.

## 7. REFERENCES

[1] Rajabhushanam C. and Kathirvel A., (2011), "Survey of Wireless MANET Application in Battlefield Operations", (IJACSA) International Journal of Advanced Computer Science and Applications, Vol. 2, No.1.

[2] www.ncs.gov

[3] IEEE Std 802.11-2007, Part 11: Wireless LAN Medium Access Control (MAC) and Physical Layer (PHY) Specifications.

[4] IEEE Standard for Information Technology – Tele communications and Information exchange between system local and metropolitan area networks – specific requirements – Part II wireless LAN medium access control(MAC) and Physical Layer(PHY) specifications, IEEE, 2007.

[5] Tariq A.S.M. and Perveen K., (2010), "Analysis of Internal Collision and Dropping Packets Characteristics of EDCA IEEE 802.11e Using NS-2.34 Simulator", Proceedings of the World Congress on Engineering and Computer Science, Vol.1, San Francisco, USA

[6] Adlen Ksentini., Abdelhak Guéroui., Mohamed Naimi., (2005), "Adaptive transmission opportunity with admission control for IEEE 802.11e networks", Proceedings of the 8th ACM international symposium on Modeling, analysis and simulation of wireless and mobile systems, Montréal, Quebec, Canada

[7] Ben Liang and Min Dong, (2007), "Packet prioritization in multihop latency aware scheduling for delay constrained communication", IEEE Journal on Selected Areas in Communications, Vol. 25, Issue: 4, pp. 819 – 830.

[8] Ku J.M, Kim S.K, Kim S.H, Simon Shin, Kim J.H and Kang C.G, (2006), "Adaptive delay threshold-based priority queuing scheme for packet scheduling in mobile broadband wireless access system", IEEE Conference on Wireless Communications and Networking.

[9] Romdhani L., Ni Q. and Turletti T., (2003), "Adaptive EDCF: enhanced service differentiation for IEEE 802.11 wireless ad hoc networks", Proceedings of the IEEE Wireless Communications and Networking (WCNC 2003), New Orleans, Louisiana, USA.

[10] Gannoune L., (2006), "A Comparative Study of Dynamic Adaptation Algorithms for Enhanced Service Differentiation in IEEE 802.11 Wireless Ad Hoc Networks", IEEE Advanced International Conference on Telecommunications and International conference on Internet and Web Applications and Services (AICT-ICIW'06).

[11] Hannah Monisha J. and Rhymend Uthariaraj V., (2012),"A Dynamic Scheduling Model for MANETs using Order Statistics", IEEE International Conference on Recent trends in Information Technology.

[12] Hannah Monisha J. and Rhymend Uthariaraj V., (2012) "User Profile based Proportional Share Scheduling and MAC protocol for MANETs.", International Journal of Distributed and Parallel Systems (IJDPS), Vol.3, No.1., pp.269-283.

[13] Kishore S. Trivedi, (2001), "Probability & Statistics with reliability, queuing, and Computer Science Applications", Prentice-Hall of India, Thirteenth Printing.

[14] Robert V. Hogg, Elliot A. Tanis and Jagan Mohan Rao, (2006), "Probability and Statistical Inference, $7^{th}$ edition", Pearson Education, Inc.

[15] Yousry H. Abdelkader and Maram Al-Wohaibi, (2011), "Computing the Performance Measures in Queuing Models via the Method of Order Statistics," Journal of Applied Mathematics, vol. 2011, Article ID 790253, 12 pages. doi:10.1155/2011/790253